\documentclass[conference]{IEEEtran}
\IEEEoverridecommandlockouts
\usepackage{cite}
\usepackage{amsmath,amssymb,amsfonts}
\usepackage{algorithmic}
\usepackage{graphicx}
\usepackage{textcomp}
\usepackage{xcolor}
\def\BibTeX{{\rm B\kern-.05em{\sc i\kern-.025em b}\kern-.08em
    T\kern-.1667em\lower.7ex\hbox{E}\kern-.125emX}}

\usepackage{epstopdf}
\usepackage{enumitem}
\usepackage{multicol}
\usepackage{multirow}
\usepackage{dsfont}
\usepackage{amsopn}

\DeclareMathOperator*{\argmin}{argmin}

\usepackage[multiple]{footmisc}

\usepackage{subcaption}
\usepackage{threeparttable}
\usepackage{float}
\usepackage{hyperref}

\usepackage{algorithm}

\newcommand{\name}{{\textsc{TintCF}}}
\newcommand{\nameA}{{\textsc{Tint-A}}}
\newcommand{\nameL}{{\textsc{Tint-L}}}
\newtheorem{problem}{Problem}

\begin{document}

\title{Time Will Tell: Harness Time to Denoise Collaborative Filtering}

\author{\IEEEauthorblockN{Xinyu He}
\IEEEauthorblockA{\textit{University of Illinois, Urbana-Champaign} \\
Champaign, IL\\
xhe34@illinois.edu}
\and
\IEEEauthorblockN{Jose Sepulveda}
\IEEEauthorblockA{\textit{Shopify}\\
Seattle, WA \\
jose.sepulveda@shopify.com}
\and
\IEEEauthorblockN{Mostafa Rahmani}
\IEEEauthorblockA{\textit{Amazon} \\
Seattle, WA \\
mostrahm@amazon.com}
\and
\IEEEauthorblockN{Alyssa Woo}
\IEEEauthorblockA{\textit{Amazon} \\
Seattle, WA \\
wooae@umich.edu}
\and
\IEEEauthorblockN{Fei Wang}
\IEEEauthorblockA{\textit{Alchedata AI} \\
Fremont, CA \\
fei.wang@alchedata.ai}
\and
\IEEEauthorblockN{Hanghang Tong}
\IEEEauthorblockA{\textit{University of Illinois, Urbana-Champaign} \\
Champaign, IL \\
htong@illinois.edu}
}

\maketitle

\begin{abstract}
    Neural graph collaborative filtering is a powerful tool for recommendation due to its superb capability of capturing the user-item interaction. 
    Despite the substantial progress, the noisy nature of interaction data remains a significant challenge in collaborative filtering, especially for data collected from implicit feedback. Both noisy interactions and interactions crafted by malicious users could greatly degenerate model performance. 
    To purify the input data and enhance model robustness, it is vital to understand how noisy interactions can be identified and rectified. Existing denoising methods focus on mining the correlation between user/item embeddings and noisy interactions, neglecting the temporal dimension for denoising. However, time information bears its natural and unique advantage in revealing noisy edges as we observe in this paper.
    In this work, we introduce a simple yet effective mechanism for generating time-aware user/item embeddings that capture both global preferences and temporal patterns. Two strategies, reweighting adjacency matrix (\nameA{}) and reweighting loss function (\nameL{}), are proposed to leverage the embeddings for the task of using \underline{t}ime \underline{in}forma\underline{t}ion to denoise \underline{c}ollaborative \underline{f}iltering (\name{}). Extensive experiments have been conducted on four real-world datasets to demonstrate the efficacy of our methods and illustrate how time information indeed helps identify and rectify noisy edges. Our codes are available at \href{https://github.com/xinyuu-he/TintCF}{https://github.com/xinyuu-he/TintCF}.
\end{abstract}


\section{Introduction}

Recommender systems have been widely applied for studying user preferences \cite{lu2015recommender}. As a powerful paradigm for building recommender systems, collaborative filtering leverages the interaction history of users and items to mine latent information. Lately, with the development of Graph Neural Networks, neural graph collaborative filtering has received extensive attention, and many powerful methods have been proposed \cite{SimGCL,NCL,lightgcn,SGL} with their capability of capturing user-item interaction. Despite the substantial progress, some critical challenges still remain, and one of them lies in the noisy nature of collected data \cite{zhang2023robust}, especially when the data is collected from implicit feedback (e.g., clicks) \cite{10.1145/3474085.3475446}. For example, users might inadvertently interact with some items (misclicks or merely out of curiosity), or some of the items might be intended for others (e.g., gifts). These noisy interactions are still considered as positive samples for model training, even though they do not align with the user's preferences. Moreover, many malicious attacks \cite{10.1145/3447548.3467233,10.1145/3274694.3274706,7824865} have been designed to craft fake interactions or users to bias recommender systems, resulting in inferior prediction results. In the context of graph neural collaborative filtering, the problem becomes even worse as message passing in graph neural networks (GNNs) will amplify the impact of noisy interactions. Therefore, identifying and rectifying noisy interactions for collaborative filtering becomes important for building robust recommender systems that align with users' true interests.

Many approaches have been proposed to tackle this issue. Depending on how samples are purified, they can be roughly divided into two categories. The first type of strategies directly remove the identified noisy samples from training data \cite{10.5555/3000375.3000390,10.1145/3474085.3475446}; while another group of works actively downweight those training samples during model training \cite{BOD,RocSE,T-CE,SGDL}. To identify noisy interactions before rectification, these works rely on either trainable (e.g., a multi-layer perceptron) or predefined functions (e.g., cosine similarity) to model the correlation between user/item embeddings and reliability.

However, none of the existing works has focused on the natural advantage of time information for denoising. The advantages of leveraging interaction time for collaborative filtering are threefold. First, interaction times can be easily collected as every implicit feedback signal has an associated timestamp that we can leverage. Second, user-side temporal patterns can help identify noisy interactions that cannot be distinguished without time information. For example, if a user usually watches horror movies at night and talk shows in the morning, a record of that user watching a horror movie in the morning is likely to be a noisy interaction, and this noisy interaction cannot be distinguished without the interaction timestamps. However, if the recommender system is aware of this temporal user pattern, it could %be available to 
recommend more talk shows instead of horror movies to the user in the morning. Third, item-side temporal patterns can help identify noisy interactions as well. For example, for some movies, the records of watching these movies mainly occur around a certain time (e.g., Christmas movies are usually viewed around Christmas). Then, a record of a user watching such movies might not align with that user's usual preferences, but given that the interaction time is close to this particular time, the model will be able to recognize this is less likely to be a noisy interaction. Beyond denoising, time information can also help give a more accurate and timely recommendation for users.

To bridge the gap between time-aware recommendation and denoising neural graph collaborative filtering, we aim to design a mechanism that can incorporate interaction time into neural graph collaborative filtering with the following desired properties: (P1) be easy to be integrated with most existing neural graph collaborative filtering models with the minimal architecture modification, (P2) be able to learn temporal patterns of users and items, and (P3) enhance the model performance in both denoising and prediction. Given that most collaborative filtering systems make the final prediction based on the dot product between user and item embeddings, we propose to generate time-aware user and item embeddings, whose dot products preserve both (1) the global preferences that previous works have been exploiting and (2) the temporal preferences modeling the probability an interaction exists at a particular timestamp, which was largely ignored by previous works. 

The proposed encoders for generating time-aware embeddings can be optimized along with the training of the base model. Furthermore, we improve two popular methods for denoising the user-item bipartite graph in recommender systems, including reweighting the adjacency matrix (\nameA{}) and reweighting the loss function (\nameL{}), by integrating our proposed time-aware embeddings into the two methods to simultaneously improve their ability to identify noisy interactions and produce more accurate predictions. \nameA{} calculates the reliability score by the similarity between time-aware user and item embeddings, then reweights and prunes edges in the user-item bipartite graph through both soft and hard assignment to remove or downweight noisy interactions. \nameL{} utilizes a weight generator that leverages time-aware embeddings to predict the probability of an interaction being noisy and minimizes the influence of noisy interactions by downweighting noisy samples in the loss function. 
We evaluate the effectiveness of proposed methods on four real-world datasets and compare them with state-of-the-art methods. We further evaluate the robustness of the proposed models against random noises. Ablation studies have  been conducted to demonstrate the effectiveness of our proposed time-aware embeddings learning mechanism. Visualization of learned weights further confirms the effectiveness of time information for detecting noisy edges.

Our main contributions are summarized as follows:
\begin{itemize}
    \item {\bf Problem.} We propose a method (\name{}) that leverages time information in neural graph collaborative filtering to enhance both denoising and prediction performance. To the best of our knowledge, this is the first work to leverage time information in denoising neural graph collaborative filtering recommender systems.
    \item {\bf Algorithms.} We propose two methods, \nameA{} and \nameL{}, that utilize the proposed time-aware embeddings for denoising recommender systems by reweighting edges or losses.
    \item {\bf Evaluations.} Extensive experiments are conducted to demonstrate efficacy of proposed methods. Our methods outperform state-of-the-art models on both vanilla and noisy datasets. 
\end{itemize}

The rest of the paper is structured as follows. Section 2 provides relevant preliminary knowledge. Section 3 introduces our proposed methods. Section 4 presents the experimental results and analysis. Section 5 discusses related works. Section 6 draws the conclusion of the paper.
\section{Preliminaries}

In this section, we first present main notations used throughout
the paper. Then we introduce the problem definition and briefly review neural graph collaborative filtering and denoising recommender system.
\begin{table}[h]
\centering
\caption{Main symbols used in the paper.}
\begin{tabular}{c | p{6.6cm}} 
\hline
Symbols & Descriptions \\ [0.5ex] 
\hline
$\mathbf{A}, v$ & a matrix, a vector \\
$\mathbf{A}^T, v^T$ & transpose of a matrix/vector\\
\hline
$\mathcal{U}, \mathcal{I}, \mathcal{T}$ & user set, item set and timestamp set \\
$\mathcal{D}$ & user-item interaction dataset \\
$\mathcal{O}$ & training set with positive and negative samples \\
$\mathcal{G}$ & bipartite graph corresponding to $\mathcal{D}$\\
$\mathbf{A}$ & adjacency matrix of $\mathcal{G}$\\
$\mathbf{E}_u, \mathbf{E}_i$ & embeddings of users and items\\
$e_u, e_i, e_t$ & embedding of a single user, item or timestamp\\
$e_u^t, e_i^t$ & time-aware embedding of a single user or item\\
$d$ & dimension of embeddings ($e_u, e_i, e_t$)\\
$d_t$ & dimension of timestamps\\
$p_{ui}$ & general preference of user $u$ over item $i$ \\
$q_{ui}^t$ & temporal preference of user $u$ over item $i$ at time $t$ \\
$p_{ui}^t$ & time-aware preference of user $u$ over item $i$ at time $t$\\
$r_{ui}^t$ & reliability score of an interaction between user $u$ and item $i$ at time $t$\\
\hline
\end{tabular}
\label{table:symbols}
\end{table}
\subsubsection{Notations}In this paper, we use caligraphic letters for sets, bold upper-case letters for two-dimensional matrices, lower-case letters for vectors or hyperparameters. We follow the conventions in PyTorch for indexing. For example, $\mathbf{A}[i,:]$ denotes the $i$-th row of matrix $\mathbf{A}$.

The user-item interaction dataset (collected from implicit feedback) is defined as $\mathcal{D}=\{(u,i,t_{ui})|u\in\mathcal{U},i\in\mathcal{I},t\in\mathcal{T}\}$, where $\mathcal{U}$ and $\mathcal{I}$ are user and item set respectively, $\mathcal{T}$ is the collection of all possible timestamps. Each timestamp is represented as a multi-dimensional vector $t\in \mathbb{R}^{d_t}$, where each dimension contains a discrete and bounded value. For example, $t$ can be in the form of [day, hour, minute, second] with $d_t=4$. The temporal bipartite graph generated from $\mathcal{D}$ is defined as $\mathcal{G}=\{\mathcal{V},\mathcal{E}\}$, where $\mathcal{V}=\mathcal{U}\cup\mathcal{I}$, $\mathcal{E}=\{(u,i)|(u,i,t_{ui})\in\mathcal{D}\}$.
Its adjacency matrix is denoted by $\mathbf{A}\in\mathbb{R}^{|\mathcal{U}|\times|\mathcal{I}|}$.

\subsubsection{Problem Definition}We formulate the temporal recommendation problem as follows.
\begin{problem}
    Temporal Recommendation.
    
    \textbf{Input}: a temporal user-item interaction dataset $\mathcal{D}$, a query from user $u$ at time $t$;

    \textbf{Output}: the prediction of item ranking according to user's preference on items at time $t$.
\end{problem}

\subsubsection{Neural Graph Collaborative Filtering}Let $f_\theta$ be a Graph Neural Network (GNN) utilized in neural graph collaborative filtering algorithm with $\theta$ as model parameters, the training of  neural graph collaborative filtering models can be formulated as,
\begin{equation}
    \theta^* = \argmin_{\theta}\mathcal{L}(\mathcal{D},\mathbf{E}_u, \mathbf{E}_i),
    \quad~(\mathbf{E}_u, \mathbf{E}_i) = f_\theta(\mathcal{D}),
\end{equation}
$\mathbf{E}_u\in\mathbb{R}^{|\mathcal{U}|\times d}, \mathbf{E}_i\in\mathbb{R}^{|\mathcal{I}|\times d}$ are learned users and items embeddings, and are the outputs of the GNN model. $\mathcal{L}$ is the loss function, $\theta^*$ is the optimal model parameter. 
Take \cite{SimGCL} as an example, $f_\theta(\cdot)$ is formulated as
\begin{align}
    &\mathbf{E}_u^{(l)} = \hat{\mathbf{A}}\mathbf{E}_i^{(l-1)},\quad\mathbf{E}_i^{(l)} = \hat{\mathbf{A}}^T\mathbf{E}_u^{(l-1)},\quad l=\{1,...,L\}\label{eq:base:emb}\\
    &\mathbf{E}_u = \frac{1}{L}\sum_{l=1}^L\mathbf{E}_u^{(l)},\quad \mathbf{E}_i = \frac{1}{L}\sum_{l=1}^L\mathbf{E}_i^{(l)},\label{eq:base:agg}
\end{align}
where $L$ is the number of layers, $\mathbf{E}_u^{(0)}$ and $\mathbf{E}_i^{(0)}$ are trainable parameters, $\hat{\mathbf{A}}=\mathbf{D}_u^{-\frac{1}{2}}\mathbf{A}\mathbf{D}_i^{-\frac{1}{2}}$ is the normalized adjacency matrix with user/item degree matrices $\mathbf{D}_u$ and $\mathbf{D}_i$ respectively.
$\mathcal{L}$ is formulated as a mixture of objectives, including BPR loss \cite{bpr}, CL loss \cite{SimGCL},
\begin{align}
    \mathcal{L}_{BPR}&=\frac{1}{|\mathcal{O}|}\sum_{(u,i,j)\in \mathcal{O}} -\log\sigma (e_u^Te_i-e_u^Te_j)
\end{align}
\begin{align}
    \mathcal{L}_{CL}&= \frac{1}{|\mathcal{U}|}\sum_{u\in\mathcal{U}} -\log\frac{\exp(\cos(e_u',e_u'')/\tau)}{\sum_{u'\in\mathcal{U}}\exp(\cos(e_u',e_{u'}'')/\tau)}\nonumber\\
    &\quad+\frac{1}{|\mathcal{I}|}\sum_{i\in\mathcal{I}} -\log\frac{\exp(\cos(e_i',e_i'')/\tau)}{\sum_{i'\in\mathcal{I}}\exp(\cos(e_i',e_{i'}'')/\tau)},
\end{align}
where $\sigma$ is Sigmoid function, $\tau >0$ is hyperparameter. $\mathcal{O}$ is the training set, each sample $(u,i,j)$ in $\mathcal{O}$ includes user $u$, positive sample item $i$ with observed interaction with $u$ in $\mathcal{D}$, and a negative sample item $j$ with no historical interaction with $u$. With time information available, each sample have additional timestamp $\mathcal{O}_t=\{(u,i,j,t_{ui})|(u,i,t_{ui})\in\mathcal{D},(u,j,t_{ui})\notin\mathcal{D}\}$. $e_j$ is the learned embedding of $j$, $e_j'$ and $e_j''$ are contrastive views of $e_j$ generated from $f_\theta(\cdot)$ with different layer-wise noises $\Delta^{(l)}$. For $l=\{1,...,L\}$,
\begin{equation}\label{eq:base:per}
    \mathbf{E}_u'^{(l)} = \hat{\mathbf{A}}\mathbf{E}_i'^{(l-1)}+\Delta_u^{(l)},\quad\mathbf{E}_i'^{(l)} = \hat{\mathbf{A}}^T\mathbf{E}_u'^{(l-1)}+\Delta_i^{(l)}.
\end{equation} 
Apart from the above objectives, AU Loss \cite{au} is another popular objective which improves the prediction performance by enforcing alignment and uniformity of representations~\cite{BOD, au} and is defined as 
\begin{align}
    \mathcal{L}_{AU} &= \frac{1}{|\mathcal{O}|}\sum_{(u,i,j)\in \mathcal{O}} \|\frac{e_u}{\|e_u\|_2}-\frac{e_i}{\|e_i\|_2}\|_2^2\nonumber\\
    &\quad+\gamma\big(\frac{1}{|\mathcal{U}|^2}\sum_{u,u'\in\mathcal{U}}\exp(-2\|\frac{e_u}{\|e_u\|_2}-\frac{e_{u'}}{\|e_{u'}\|_2}\|_2^2)\nonumber\\
    &\quad+\frac{1}{|\mathcal{I}|^2}\sum_{i,i'\in\mathcal{I}}\exp(-2\|\frac{e_i}{\|e_i\|_2}-\frac{e_{i'}}{\|e_{i'}\|_2}\|_2^2)\big).
\end{align}
The first term is the alignment loss which enforces distance between normalized embeddings of positive pairs to be minimized. The second and third terms are uniformity losses in user embeddings space and item embeddings space, distributing normalized embeddings uniformly on the unit hypersphere. $\gamma$ is a hyperparameter to balance alignment and uniformity. 

Finally, dot products of embeddings are applied to measure the preference of user $u$ towards item $i$,
\begin{equation}
    p_{ui} = e_u^Te_i, \forall u\in\mathcal{U}, i\in\mathcal{I}
\end{equation}

\subsubsection{Denoising recommender systems}
Neural graph collaborative filtering has been proved to be successful in many previous works \cite{NCL, SimGCL, lightgcn} whose effectiveness is fundamentally driven by the quality of the input training data. Therefore, these methods might be negatively impacted by the noisy interactions in the collected data. To solve this problem, many solutions for rectifying noisy interactions have been proposed. One common way is to reweight or delete the edges of interactions in $\mathcal{G}$ \cite{RocSE,RGCF}. With a reliability score function $r_{ui}=\textrm{R}(e_u, e_i)$ that takes user and item embeddings as input, denoising models try to reweight and selectively delete edges by the reliability of observed interactions.
\begin{equation}\label{eq:de:adj}
    \Tilde{\mathbf A}[u,i] = \mathds{1}(r_{ui}>\beta)r_{ui},\forall (u,i)\in\mathcal{E}
\end{equation}
where $\Tilde{\mathbf{A}}$ is the adjacency matrix of the reweighted graph, $\mathds{1}(\cdot)$ is indicator function and $\beta$ is a hyperparameter. In this way, edges with lower reliability scores will be deleted and interactions with higher reliability scores will be assigned higher weights in $\Tilde{\mathbf{A}}$. With the reweighted adjacency matrix, noisy interactions will take less or no effect in message passing in GNNs.

Likewise, another common approach \cite{BOD, T-CE} rectifies noisy interactions by reweighting loss function with reliability of interactions. For example, with a weight generation function $w_{ui} = \textrm{W}(e_u, e_i)$, the reweighted BPR loss will be formulated as
\begin{equation}
    \mathcal{L}_{BPR}=\frac{1}{|\mathcal{O}|}\sum_{(u,i,j)\in \mathcal{O}} -\log\sigma (w_{ui}e_u^Te_i-w_{uj}e_u^Te_j).
\end{equation}
This approach can also be applied to CL loss and alignment loss.
Note that the model also reweights negative samples, where the weight generation module $\textrm{W}(e_u, e_i)$ is assumed to be capable of automatically learning to first distinguish positive samples and negative samples, and then output how reliable the positive (or negative) interaction is. 

\begin{algorithm}[!h]
\caption{\nameA{}: Denoise Collaborative Filtering by Reweighting the Adjacency Matrix}
\begin{algorithmic}[1]
\REQUIRE (1) the dataset $\mathcal{D}$, and (2) hyperparameters: threshold $\beta$, weights of CL loss and AU loss $\lambda_1, \lambda_2$; 
    %\STATEx Dataset $\mathcal{D}$, hyperparameters: $\beta, \lambda_1, \lambda_2$
\ENSURE
    (1) user/item embeddings $\mathbf{E}_u, \mathbf{E}_i$; and (2) optimal parameters in timestamp encoder $\{\mathbf{W}_i\}_{i=1,...,d_t}$.
    \STATE Randomly initialize $\mathbf{E}_u^{(0)}, \mathbf{E}_i^{(0)}, \{\mathbf{W}_i\}_{i=1,...,d_t}$\;
    \WHILE{not converged}
        \STATE Derive $e_t$ for all timestamps in $\mathcal{D}$ with Eqs.~\eqref{eq:time:st}--\eqref{eq:time:end}\;
        \STATE Construct reweighted adjacency matrix with Eqs.~\eqref{eq:de:r} and Eq.~\eqref{eq:de:adj}\;
        \STATE Derive $\mathbf{E}_u, \mathbf{E}_i$ from Eqs.~\eqref{eq:base:emb:r} and \eqref{eq:base:agg}\;
        \STATE Derive $\mathbf{E}'_u, \mathbf{E}'_i$ and $\mathbf{E}''_u, \mathbf{E}''_i$ from Eqs.~\eqref{eq:base:per:r} and \eqref{eq:base:agg}\;
        \STATE Sample a batch of training samples $(u,i,j,t_{ui})$
        \STATE Derive loss $\mathcal{L}$ of the batch with Eq.~\eqref{eq:loss}\;
        \STATE Update $\mathbf{E}_u^{(0)}, \mathbf{E}_i^{(0)}, \{\mathbf{W}_i\}_{i=1,...,d_t}$ with gradient descent\;
    \ENDWHILE
\end{algorithmic}
\label{alg:code:A}
\end{algorithm}

\begin{algorithm}[!h]
\caption{\nameL{}: Denoise Collaborative Filtering by Reweighting the Loss Function}
\begin{algorithmic}[1]
\REQUIRE
    (1) the dataset $\mathcal{D}$, and (2) hyperparameters: weights of CL loss and AU loss $\lambda_1, \lambda_2$;
\ENSURE
    (1) user/item embeddings $\mathbf{E}_u, \mathbf{E}_i$; and (2) optimal parameters in timestamp encoder $\{\mathbf{W}_i\}_{i=1,...,d_t}$.
    \STATE Randomly initialize $\mathbf{E}_u^{(0)}, \mathbf{E}_i^{(0)}, \{\mathbf{W}_i\}_{i=1,...,d_t}$, and parameters in weight generator $\textrm{W}(\cdot)$\;
    \WHILE{not converged}
        \STATE Sample a batch of training samples $\mathcal{B}=(u,i,j,t_{ui})$
        \STATE Derive $e_t$ for all $t_{ui}\in\mathcal{B}$ with Eqs.~\eqref{eq:time:st}--\eqref{eq:time:end}\;
        \STATE Generate $w_{ui}, w_{uj}$ for all samples in $\mathcal{B}$\;
        \STATE Derive $\mathbf{E}_u, \mathbf{E}_i$ from Eqs.~\eqref{eq:base:emb} and \eqref{eq:base:agg}\;
        \STATE Derive $\mathbf{E}'_u, \mathbf{E}'_i$ and $\mathbf{E}''_u, \mathbf{E}''_i$ from Eqs.~\eqref{eq:base:per} and \eqref{eq:base:agg}\;
        \STATE Derive loss $\mathcal{L}$ with reweighted version of Eq.~\eqref{eq:loss}\;
        \STATE Update $\mathbf{E}_u^{(0)}, \mathbf{E}_i^{(0)}, \{\mathbf{W}_i\}_{i=1,...,d_t}$ with gradient descent\;
        \STATE Record $\nabla_{\theta}\mathcal{L}_{BPR}, \nabla_{\theta}\mathcal{L}_{AU}$\;
        \STATE Derive loss $\mathcal{L}_{GM}$ in Eq.~\eqref{eq:loss:gm}\;
        \STATE Update $\textrm{W}(\cdot)$ with gradient descent\;
    \ENDWHILE
\end{algorithmic}
\label{alg:code:L}
\end{algorithm}

\section{Proposed Framework: \name{}}
To maximally improve the model's performance in both prediction and denoising, we aim to find an effective and efficient way to incorporate time information into neural graph collaborative filtering frameworks by defining time-aware embeddings. Here, a key challenge is to ensure the dot product of embeddings to remain meaningful for preference prediction. 
In this section, we present our proposed method for yielding time-aware embeddings (Section 3.1) and how to integrate with two denoising methods (Sections 3.2 and 3.3).
\subsection{Time-aware Embeddings}
Most existing methods rely on embeddings to decide reliability or preferences. Hence, it is natural to define time-aware embeddings for boosting existing works' performance. Here, the key challenge lies in how to ensure the time-aware embeddings' dot products preserve the user preference while reflecting temporal patterns as well. Therefore, we design the dot product so that it can be decomposed into two parts: $p_{ui}$ that represents user's general preference of item $i$, and $q_{ui}^t$ referring to user's temporal preference of item $i$ at time $t$. Then we can define an overall time-aware preference $p_{ui}^t$ of user $u$ over item $i$ at time $t$ as %\hh{let's put some of these notations in Table 1. this helps with the readability}
\begin{equation}
    p_{ui}^t = p_{ui} + q_{ui}^t = e_u^Te_i + q_{ui}^t.
\end{equation}
To fit the dot product paradigm in existing neural graph collaborative filtering frameworks, we seek a definition of $q_{ui}^t$ so that $p_{ui}^t$ can be decomposed into dot product of users' and items' temporal embeddings as well
\begin{equation}\label{eq:def:p}
    p_{ui}^t=(e_u^t)^Te_i^t = e_u^Te_i + q_{ui}^t
\end{equation}
where $e_u^t, e_i^t$ are time-aware embeddings of user $u$ and item $i$ that we need to define.
To start with, we expect $q_{ui}^t$ to include a dot product term that estimate the similarity between an embedding of one interaction $e_{ui}$ and a timestamp embedding that concludes the overall user-item temporal pattern at time $t$,
\begin{equation}\label{eq:def:mid:quit}
    q_{ui}^t = (e_{ui})^Te_t
\end{equation}
where $e_{ui}$ represents the joint embedding of user $u$ and item $i$ which can be defined through addition $e_{ui}=e_u+e_i$, and $e_t$ captures the temporal trends at time $t$. However, solving $e_u^t$ and $e_i^t$ in Eq.~\eqref{eq:def:p} is highly non-trivial, and embeddings instead of preferences are required for some loss functions (e.g., CL loss and AU loss) or weight generation functions (e.g., \cite{BOD, RocSE, RGCF}). To resolve this issue, we redefine $q_{ui}^t$ as
\begin{equation}\label{eq:def:quit}
    q_{ui}^t = (e_{ui})^Te_t + \|e_t\|_2^2.
\end{equation}
Note that at a specific time $t$, $\|e_t\|_2^2$ is the same for all user-item pairs. Therefore, the additional term in Eq.~\eqref{eq:def:quit} compared with Eq.~\eqref{eq:def:mid:quit} will not affect the ranking of items. With the definition of $q_{ui}^t$ in Eq.~\eqref{eq:def:quit}, time-aware preference can be rewritten as 
\begin{align}\label{eq:puit}
    p_{ui}^t &= (e_u^t)^Te_i^t = e_u^Te_i + q_{ui}^t = e_u^Te_i + (e_{ui})^Te_t + \|e_t\|_2^2 \nonumber\\
    &= e_u^Te_i + e_u^Te_t+e_i^Te_t+\|e_t\|_2^2 = (e_u+e_t)^T(e_i+e_t)
\end{align}
and a natural choice for $e_u^t$ and $e_i^t$ is
\begin{equation}
    e_u^t = e_u+e_t,\quad e_i^t = e_i+e_t.
\end{equation}
With Eq.~\eqref{eq:puit}, $p_{ui}^t$ is capable of capturing both user-side and item-side temporal patterns. To capture item-side temporal patterns, it is reasonable to assume that trained time embeddings can capture item temporal distribution. For example, we expect the embedding of a timestamp $t$ near Christmas to be close to Christmas movies’ embeddings. In this way, $e_i^Te_t$ will be positive for Christmas movies and users will likely be recommended with Christmas movies. Similarly, to capture user-side temporal patterns, we expect the embedding of a timestamp to be close to the summation of user-item embeddings that tend to occur at this specific time.

Finally, we design an encoder of timestamps to learn $e_t$. The idea of incorporating timestamps have been studied in sequential models \cite{10.1145/3604915.3608830,TGSRec,kazemi2019time2vec}, and we use a simplified variant of \cite{10.1145/3604915.3608830} to learn timestamp embeddings. For a timestamp $t\in\mathcal{T}$ where each dimension is a discrete value (e.g., representing [day, hour, minute, second]), each dimension will be first encoded into a one-hot vector and further embedded with a learnable matrix. Then embeddings of each dimension will be concatenated into the embedding of timestamp. Define learnable matrices as $\{\mathbf{W}_{i}\}_{i=1,2,...,d_t}$, the embedding of timestamp $t$ will be
\begin{align}
    e_t^{(i)} &= \operatorname{one-hot}(t[i])\mathbf{W}_i,\quad i=1,2,...,d_t,\label{eq:time:st}\\
    e_t &= [e_t^{(1)} \| e_t^{(2)}\|...\|e_t^{(d_t)}],\\
    \mathbf{W}_{i}&\in\mathbb{R}^{m_i\times(\lfloor d/d_t\rfloor+\mathds{1}(i\leq d\%d_t))},\label{eq:time:end}
\end{align}
where $m_i$ is the number of distinct values in dimension $i$ for all timestamps in $\mathcal{T}$, $\|$ denotes concatenation, $\%$ stands for modulo operator. $\mathds{1}(i\leq d\%d_t)$ is added to align the dimension of $e_t$ with the dimensions of $e_u$ and $e_i$. The space complexity for storing $\{\mathbf{W}_{i}\}_{i=1,2,...,d_t}$ is $O((\sum_{i=1}^{d_t}m_i) d/d_t)$, while the time complexity for calculating $T$ timestamp embeddings is $O(Td)$.

A major advantage of the proposed time-aware embeddings lies in that it can be naturally integrated with various denoising strategies in the  neural graph collaborative filtering models. In the next two subsections, we present two instantiations, including reweighting the adjacency matrix and reweighting the loss function.

\subsection{\nameA{}: Reweighting the Adjacency Matrix}
We define the cosine similarity based reliability score function with time-aware embeddings as 
\begin{equation}\label{eq:de:r}
    r^t_{ui} = (\cos(e_u^{(0)}+e_t,e_i^{(0)}+e_t)+1)/2.
\end{equation}
%Note that we differ from \cite{RocSE}: \cite{RocSE} uses the second and first layer embeddings $e_u^{(2)}, e_i^{(1)}$, while we use the zero-th layer embeddings. 
Take \cite{SimGCL} as the backbone model, the time-aware neural graph collaborative filtering process is formulated as %\hh{$E'$ instead of $E^'$, same change in other places}
\begin{align}
    &\mathbf{E}_u'^{(l)} = \Tilde{\mathbf{A}}\mathbf{E}_i'^{(l-1)}+\Delta_u^{(l)},\quad\mathbf{E}_i'^{(l)} = \Tilde{\mathbf{A}}^T\mathbf{E}_u'^{(l-1)}+\Delta_i^{(l)},\label{eq:base:per:r} \\
    &\mathbf{E}_u^{(l)} = \Tilde{\mathbf{A}}\mathbf{E}_i^{(l-1)},\quad\mathbf{E}_i^{(l)} = \Tilde{\mathbf{A}}^T\mathbf{E}_u^{(l-1)},\quad l=\{1,...,L\}. \label{eq:base:emb:r}
    % &p^t_{ui} = (e_u+e_t)^T(e_i+e_t),  \forall u\in\mathcal{U}, i\in\mathcal{I}, t\in\mathcal{T}\label{eq:pre:time}
\end{align}
where $\Tilde{\mathbf A}$ is the reweighted adjacency matrix in Eq.~\eqref{eq:de:adj}. We apply the noise sampling strategy that is shown to be effective in \cite{RocSE}, %highly effective for denoising model with the reweighted adjacency matrix in \cite{RocSE},
\begin{align}
    \mathbf{Z}_u^{(l)} &= \operatorname{Shuffle}(\hat{\mathbf{E}}_u^{(l)}), \quad
    \Delta_u^{(l)} = \epsilon \mathbf{Z}_u^{(l)},\nonumber\\
    \mathbf{Z}_i^{(l)} &= \operatorname{Shuffle}(\hat{\mathbf{E}}_i^{(l)}), \quad
    \Delta_i^{(l)} = \epsilon \mathbf{Z}_i^{(l)},\quad l=\{1,...,L\},
\end{align}
where $\hat{\mathbf{E}}_u^{(l)}, \hat{\mathbf{E}}_i^{(l)}$ are row-normalized $\mathbf{E}_u^{(l)}, \mathbf{E}_u^{(l)}$, and $\operatorname{Shuffle}(\mathbf{X})$ shuffles the rows in $\mathbf{X}$. In this noise sampling strategy, noises are sampled from the embedding space and this helps mimic real graph structural attacks. Finally, we jointly train the timestamp encoder and parameters of collaborative filtering model ($\mathbf{E}_u^{(0)}, \mathbf{E}_u^{(0)}$) with a combination of BPR, CL and AU loss
\begin{equation}\label{eq:loss}
    \mathcal{L} = \mathcal{L}_{BPR}+\lambda_1\mathcal{L}_{CL}+\lambda_2\mathcal{L}_{AU}
\end{equation}
where $\lambda_1, \lambda_2$ are weights of CL loss and AU loss. %Compared to the loss function in \cite{SimGCL}, we replace $L_2$ regularization term with the AU loss owing to its effectiveness in denoising recommender systems \cite{BOD}. 

Although \cite{RocSE} also uses a cosine similarity based reliability score for denoising neural graph collaborative filtering, the proposed \nameA{} bears two distinct advantages. First (computation), \cite{RocSE} relies on the second and first layer embeddings $e_u^{(2)}, e_i^{(1)}$, which doubles the computation in both forward and backward processes. %Therefore, to reduce the computational cost in reliability score calculation, 
In contrast, we use the zero-th layer embeddings in Eq.~\eqref{eq:de:r} and incorporate \name{} embeddings to improve model utility and robustness. Second (effectiveness), the objective in Eq.~\eqref{eq:loss} differs from \cite{RocSE}: we replace $L_2$ regularization term with the AU loss owing to its effectiveness in recommender systems \cite{SimGCL, BOD}. In addition, uniformity loss is applied to randomly sampled negative user-item pairs instead of positive pairs to avoid the popularity bias.

Pseudo codes of \nameA{} shown in Algorithm.\ref{alg:code:A}.

\begin{table*}[t]
    \centering
    \caption{Experimental results on vanilla datasets. The best results are in bold. The best results in baselines are underlined. Stds(\%) are in footnotes.}
    \resizebox{\textwidth}{!}{
    \begin{tabular}{c|c|c|c|c|c|c|c|c|c|c|c}
        \hline
        Type & Dataset & Metric & TGSRec & TIN & T-CE & DeCA & DDRM & BOD & PLD & \nameA{} & \nameL{} \\
        \hline
        \multirow{12}*{Explicit} & \multirow{6}*{ML-100K}
         & Prec@10  & $0.1400_{0.59}$ & $0.1379_{0.48}$ & $0.1581_{1.26}$ & $0.1959_{0.54}$ & $0.1640_{1.73}$ & $\underline{0.2197}_{0.29}$ & $0.1869_{0.11}$ & $\textbf{0.2373}_{0.96}$ & $0.2188_{0.45}$ \\
         & & Recall@10 & $0.0523_{0.29}$ & $0.0666_{0.23}$ & $0.0712_{0.65}$ & $0.0878_{0.32}$ & $0.0712_{0.81}$ & $\underline{0.0949}_{0.19}$ & $0.0824_{0.07}$ & $\textbf{0.1026}_{0.72}$ & $0.0959_{0.17}$ \\
         & & NDCG@10 & $0.1461_{0.81}$ & $0.1161_{0.37}$ & $0.1662_{1.40}$ & $0.2053_{0.67}$ & $0.1737_{1.80}$ & $\underline{0.2652}_{0.39}$ & $0.1081_{0.05}$ & $0.2542_{1.05}$ & $\textbf{0.2672}_{0.31}$ \\
         & & Prec@20 & $0.1236_{0.29}$ & $0.1311_{0.37}$ & $0.1438_{1.04}$ & $0.1737_{0.38}$ & $0.1452_{1.13}$ & $\underline{0.1955}_{0.13}$ & $0.1622_{0.09}$ & $\textbf{0.2063}_{0.41}$ & $0.1974_{0.23}$ \\
         & & Recall@20 & $0.0916_{0.24}$ & $0.1234_{0.25}$ & $0.1266_{0.90}$ & $0.1510_{0.36}$ & $0.1244_{1.01}$ & $\underline{0.1678}_{0.20}$ & $0.1414_{0.07}$ & $\textbf{0.1746}_{0.66}$ & $0.1699_{0.28}$ \\
         & & NDCG@20 & $0.1438_{0.56}$ & $0.1181_{0.34}$ & $0.1730_{1.38}$ & $0.2096_{0.52}$ & $0.1753_{1.50}$ & $\underline{0.2692}_{0.31}$ & $0.1487_{0.04}$ & $0.2532_{0.74}$ & $\textbf{0.2734}_{0.23}$ \\
        \cline{2-12}
        & \multirow{6}*{ML-1M}
         & Prec@10 & $0.1150_{0.62}$ & $0.1063_{0.03}$ & $0.1545_{0.16}$ & ${0.1635}_{0.08}$ & $0.1132_{0.90}$ & $0.1422_{0.28}$ & $\textbf{0.1792}_{0.05}$ & $0.1765_{0.05}$ & $0.1703_{0.25}$ \\
         & & Recall@10 & $0.0287_{0.18}$ & $0.0372_{0.03}$ & $0.0464_{0.05}$ & $0.0480_{0.10}$ & $0.0306_{0.29}$ & $0.0377_{0.02}$ & $\textbf{0.0553}_{0.01}$ & $0.0503_{0.04}$ & $0.0491_{0.08}$ \\
         & & NDCG@10 & $0.1219_{0.48}$ & $0.0892_{0.05}$ & $0.1624_{0.18}$ & $0.1728_{0.10}$ & $0.1165_{1.03}$ & $\textbf{0.1889}_{0.14}$ & $0.0783_{0.02}$ & $0.1872_{0.02}$ & $0.1848_{0.38}$ \\
         & & Prec@20 & $0.1048_{0.30}$  & $0.1048_{0.07}$ & $0.1404_{0.10}$ & $0.1468_{0.13}$ & $0.1002_{0.59}$ & $0.1316_{0.33}$ & $\textbf{0.1629}_{0.05}$ & $0.1576_{0.19}$ & $0.1508_{0.19}$ \\
         & & Recall@20 & $0.0511_{0.32}$ & $0.0718_{0.04}$ & $0.0822_{0.09}$ & $0.0845_{0.21}$ & $0.0529_{0.42}$ & $0.0676_{0.13}$ & $\textbf{0.0988}_{0.01}$ & $0.0869_{0.07}$ & $0.0842_{0.14}$ \\
         & & NDCG@20 & $0.1169_{0.35}$ & $0.0935_{0.06}$ & $0.1585_{0.14}$ & $0.1669_{0.09}$ & $0.1113_{0.8}$ & $\textbf{0.1805}_{0.14}$ & $0.1112_{0.03}$ & $0.1795_{0.13}$ & $0.1777_{0.31}$ \\
        \hline
        \multirow{12}*{Implicit} & \multirow{6}*{Tmall}
         & Prec@10  & -- OOM -- & $0.0008_{0.01}$ & $0.0004_{0.00}$ & $0.0010_{0.00}$ & $0.0022_{0.03}$ & $0.0058_{0.04}$ & $\underline{0.0074}_{0.01}$ & $\textbf{0.0088}_{0.01}$ & $0.0078_{0.01}$ \\
         & & Recall@10 & -- OOM -- & $0.0040_{0.07}$ & $0.0023_{0.03}$ & $0.0057_{0.02}$ & $0.0163_{0.18}$ & $0.0399_{0.28}$ & $\underline{0.0559}_{0.04}$ & $\textbf{0.0643}_{0.07}$ & $0.0557_{0.03}$ \\
         & & NDCG@10  & -- OOM -- & $0.0007_{0.01}$ & $0.0017_{0.01}$ & $0.0038_{0.00}$ & $0.0114_{0.12}$ & $0.0235_{0.19}$ & $\underline{0.0331}_{0.02}$ & $\textbf{0.0372}_{0.01}$ & $0.0326_{0.03}$ \\
         & & Prec@20 & -- OOM --  & $0.0006_{0.00}$ & $0.0003_{0.00}$ & $0.0007_{0.00}$ & $0.0014_{0.01}$ & $0.0046_{0.02}$ & $\underline{0.0050}_{0.01}$ & $\textbf{0.0064}_{0.01}$ & $0.0059_{0.00}$ \\
         & & Recall@20 & -- OOM -- & $0.0072_{0.03}$ & $0.0036_{0.03}$ & $0.0078_{0.03}$ & $0.0201_{0.22}$ & $0.0637_{0.31}$ & $\underline{0.0754}_{0.05}$ & $\textbf{0.0936}_{0.07}$ & $0.0850_{0.05}$ \\
         & & NDCG@20 & -- OOM --  & $0.0006_{0.00}$ & $0.0020_{0.02}$ & $0.0045_{0.01}$ & $0.0123_{0.13}$ & $0.0300_{0.19}$ & $\underline{0.0384}_{0.02}$ & $\textbf{0.0452}_{0.01}$ & $0.0405_{0.03}$ \\
        \cline{2-12}
        & \multirow{6}*{Music4all}
         & Prec@10 & -- OOM --  & -- OOM -- & $0.0277_{0.00}$ & $0.0362_{0.21}$ & $0.0196_{0.19}$ & $0.0558_{0.56}$ & $\underline{0.0594}_{0.06}$ & $0.0601_{0.02}$ & $\textbf{0.0636}_{0.11}$ \\
         & & Recall@10 & -- OOM -- & -- OOM -- & $0.0056_{0.00}$ & $0.0082_{0.05}$ & $0.0104_{0.04}$ & $0.0160_{0.23}$ & $\textbf{0.0201}_{0.01}$ & $0.0183_{0.03}$ & $0.0194_{0.05}$ \\
         & & NDCG@10 & -- OOM --  & -- OOM -- & $0.0295_{0.00}$ & $0.0388_{0.21}$ & $0.0244_{0.19}$ & $\underline{0.0628}_{0.66}$ & $0.0262_{0.02}$ & $0.0670_{0.02}$ & $\textbf{0.0723}_{0.09}$ \\
         & & Prec@20 & -- OOM --  & -- OOM -- & $0.0251_{0.00}$ & $0.0321_{0.18}$ & $0.0161_{0.15}$ & $0.0480_{0.48}$ & $\underline{0.0515}_{0.05}$ & $0.0515_{0.02}$ & $\textbf{0.0545}_{0.09}$ \\
         & & Recall@20 & -- OOM -- & -- OOM -- & $0.0102_{0.00}$ & $0.0143_{0.10}$ & $0.0141_{0.07}$ & $0.0262_{0.39}$ & $\textbf{0.0323}_{0.04}$ & $0.0298_{0.01}$ & $0.0316_{0.08}$ \\
         & & NDCG@20 & -- OOM --  & -- OOM -- & $0.0278_{0.00}$ & $0.0363_{0.20}$ & $0.0230_{0.17}$ & $\underline{0.0581}_{0.63}$ & $0.0351_{0.03}$ & $0.0622_{0.01}$ & $\textbf{0.0670}_{0.10}$ \\
        \hline
    \end{tabular}}
    \label{exp:vanilla}
\end{table*}

\subsection{\nameL{}: Reweighting the Loss Function}

%\hh{same comment as the above: elaborate the key difference between tint-L and 27}
In this section, we present how to integrate our time-aware embeddings with another type of denoising method by reweighting the loss function. We define the time-aware weight generator as
\begin{equation}
    w_{ui} = \textrm{W}(e_u\|e_i\|e_{t_{ui}}),
\end{equation}
where $\textrm{W}(\cdot)$ is implemented as a 3-layer Multi-layer Perceptron (MLP). The backbone model is chosen as \cite{SimGCL}. The backbone model and weight generator $\textrm{W}(\cdot)$ are alternatively updated during the whole training process.
% Same as \nameA{}, the loss function for training backbone model is formulated as a weighted combination of BPR, CL and AU loss, only that the loss terms are reweighted by $w_{ui}$. For example, the reweighted BPR loss with time-aware embeddings is
% \begin{align}
%     \mathcal{L}_{BPR}=\frac{1}{|\mathcal{O}|}\sum_{(u,i,j, t_{ui})\in \mathcal{O}} &-\log\sigma (w_{ui}(e_u+e_{t_{ui}})^T(e_i+e_{t_{ui}})\nonumber\\
%     &\quad-w_{uj}(e_u+e_{t_{ui}})^T(e_j+e_{t_{ui}})).
% \end{align}
Specifically, the time-aware weight generator $\textrm{W}(\cdot)$ is trained with a gradient matching objective \cite{BOD}, while the base model and timestamp encoder are jointly updated with reweighted Eq.~\eqref{eq:loss}. The gradient matching objective aligns the gradient of the BPR loss with the gradient of the AU loss with respect to backbone model parameters:
\begin{align}
    &\theta = [\mathbf{E}_u^{(0)}\|\mathbf{E}_i^{(0)}]\in\mathbb{R}^{(|\mathcal{U}|+|\mathcal{I}|)\times d}, \\
    &\mathbf{G}_1 = \nabla_{\theta}\mathcal{L}_{BPR}, \mathbf{G}_2 = \nabla_{\theta}\mathcal{L}_{AU},\\
    &\mathcal{L}_{GM} = \sum_{i=1}^{|\mathcal{U}|+|\mathcal{I}|}(1-\frac{\mathbf{G}_1[i,:]^T\mathbf{G}_2[i,:]}{\|\mathbf{G}_1[i,:]\|_2\|\mathbf{G}_2[i,:]\|_2}).\label{eq:loss:gm}
\end{align}
where both $\mathcal{L}_{BPR}$, $\mathcal{L}_{AU}$ are reweighted objectives.

Compared with \cite{BOD}, our time-aware weight generator enhances the capability of denoising module to determine noisy edges with the help of time information. Furthermore, the plug-and-play fashion of time-aware weight generator in \nameL{} ensures its scalability. If the hidden dimension in MLP is $d_0$, the additional time complexity in the forward process is only $O(|\mathcal{B}|(d+dd_0))$, where $|\mathcal{B}|$ is the number of positive and negative edges sampled in each batch.

Pseudo codes of \nameL{} shown in Algorithm.\ref{alg:code:L}.

\begin{table*}[t]
    \centering
    \caption{Experimental results on noisy datasets with 20\% noisy interactions. The best results are in bold. The best results in baselines are underlined. Stds(\%) are in footnotes.}
    \resizebox{\textwidth}{!}{
    \begin{tabular}{c|c|c|c|c|c|c|c|c|c|c|c}
        \hline
        Type & Dataset & Metric & TGSRec & TIN & T-CE & DeCA & DDRM & BOD & PLD & \nameA{} & \nameL{} \\
        \hline
        \multirow{12}*{Explicit} & \multirow{6}*{ML-100K}
         & Prec@10 & $0.0541_{4.23}$ & $0.1328_{0.12}$ & $0.1474_{0.11}$ & $0.1948_{0.45}$ & $0.1511_{1.52}$ & $\underline{0.2140}_{0.27}$ & $0.1860_{0.13}$ & $\textbf{0.2428}_{0.50}$ & $0.2211_{0.63}$ \\
         & & Recall@10 & $0.0174_{1.84}$ & $0.0643_{0.32}$ & $0.0676_{0.16}$ & $0.0858_{0.17}$ & $0.0654_{0.69}$ & $\underline{0.0944}_{0.04}$ & $0.0822_{0.05}$ & $\textbf{0.1060}_{0.17}$ & $0.0954_{0.37}$ \\
         & & NDCG@10 & $0.0534_{3.98}$  & $0.1125_{0.11}$ & $0.1542_{0.07}$ & $0.2056_{0.44}$ & $0.1623_{1.83}$ & $\underline{0.2618}_{0.56}$ & $0.1077_{0.06}$ & $0.2607_{0.59}$ & $\textbf{0.2666}_{0.51}$ \\
         & & Prec@20 & $0.0500_{3.44}$ & $0.1268_{0.07}$ & $0.1308_{0.07}$ & $0.1761_{0.34}$ & $0.1333_{1.32}$ & $\underline{0.1916}_{0.33}$ & $0.1622_{0.22}$ & $\textbf{0.2086}_{0.29}$ & $0.1958_{0.34}$ \\
         & & Recall@20 & $0.0309_{2.82}$ & $0.1186_{0.37}$ & $0.1193_{0.01}$ & $0.1528_{0.30}$ & $0.1141_{1.27}$ & $\underline{0.1665}_{0.18}$ & $0.1418_{0.13}$ & $\textbf{0.1763}_{0.04}$ & $0.1687_{0.46}$ \\
         & & NDCG@20 & $0.0541_{3.87}$ & $0.1145_{0.02}$ & $0.1589_{0.03}$ & $0.2123_{0.41}$ & $0.1624_{1.82}$ & $\underline{0.2674}_{0.62}$ & $0.1488_{0.13}$ & $0.2576_{0.37}$ & $\textbf{0.2706}_{0.41}$ \\
        \cline{2-12}
        & \multirow{6}*{ML-1M}
         & Prec@10 & $0.0325_{0.94}$ & $0.0730_{0.13}$ & $0.1429_{0.59}$ & $0.1609_{0.30}$ & $0.0969_{0.97}$ & $0.1472_{0.61}$ & $\textbf{0.1805}_{0.05}$ & $0.1691_{0.11}$ & $0.1719_{0.31}$ \\
         & & Recall@10 & $0.0065_{0.20}$ & $0.0255_{0.15}$ & $0.0428_{0.13}$ & $0.0474_{0.12}$ & $0.0241_{0.29}$ & $0.0384_{0.28}$ & $\textbf{0.0564}_{0.02}$ & $0.0443_{0.04}$ & $0.0504_{0.18}$ \\
         & & NDCG@10 & $0.0310_{0.71}$ & $0.0614_{0.07}$ & $0.1503_{0.64}$ & $0.1698_{0.26}$ & $0.0993_{1.22}$ & $\underline{0.1852}_{0.56}$ & $0.0793_{0.02}$ & $0.1798_{0.10}$ & $\textbf{0.1859}_{0.35}$ \\
         & & Prec@20 & $0.0315_{0.90}$ & $0.0733_{0.16}$ & $0.1291_{0.55}$ & $0.1458_{0.16}$ & $0.0823_{0.38}$ & $0.1354_{0.48}$ & $\textbf{0.1648}_{0.07}$ & $0.1519_{0.14}$ & $0.1515_{0.27}$ \\
         & & Recall@20 & $0.0126_{0.42}$ & $0.0507_{0.21}$ & $0.0761_{0.24}$ & $0.0845_{0.14}$ & $0.0403_{0.33}$ & $0.0689_{0.44}$ & $\textbf{0.1011}_{0.05}$ & $0.0782_{0.17}$ & $0.0860_{0.36}$ \\
         & & NDCG@20 & $0.0317_{0.79}$ & $0.0652_{0.10}$ & $0.1461_{0.62}$ & $0.1652_{0.18}$ & $0.0917_{0.75}$ & $\underline{0.1769}_{0.58}$ & $0.1130_{0.04}$ & $0.1718_{0.15}$ & $\textbf{0.1781}_{0.34}$ \\
        \hline
        
        % ===================== Tmall =====================
        \multirow{12}*{Implicit} & \multirow{6}*{Tmall}
         & Prec@10  & -- OOM --   & $0.0010_{0.01}$ & $0.0004_{0.00}$ & $0.0007_{0.00}$ & $0.0015_{0.01}$ & $0.0052_{0.03}$ & $\underline{0.0064}_{0.00}$ & $\textbf{0.0080}_{0.01}$ & $0.0069_{0.01}$ \\
         & & Recall@10 & -- OOM --  & $0.0063_{0.05}$ & $0.0025_{0.00}$ & $0.0039_{0.03}$ & $0.0110_{0.06}$ & $0.0352_{0.25}$ & $\underline{0.0481}_{0.05}$ & $\textbf{0.0570}_{0.07}$ & $0.0479_{0.08}$ \\
         & & NDCG@10  & -- OOM --   & $0.0009_{0.01}$ & $0.0016_{0.00}$ & $0.0026_{0.02}$ & $0.0078_{0.04}$ & $0.0208_{0.13}$ & $\underline{0.0284}_{0.03}$ & $\textbf{0.0331}_{0.02}$ & $0.0281_{0.03}$ \\
         & & Prec@20   & -- OOM --  & $0.0008_{0.00}$ & $0.0003_{0.00}$ & $0.0005_{0.00}$ & $0.0009_{0.00}$ & $0.0040_{0.02}$ & $\underline{0.0043}_{0.01}$ & $\textbf{0.0057}_{0.00}$ & $0.0051_{0.01}$ \\
         & & Recall@20 & -- OOM --  & $0.0092_{0.01}$ & $0.0034_{0.00}$ & $0.0052_{0.02}$ & $0.0132_{0.08}$ & $0.0548_{0.32}$ & $\underline{0.0645}_{0.07}$ & $\textbf{0.0820}_{0.08}$ & $0.0707_{0.14}$ \\
         & & NDCG@20  & -- OOM --   & $0.0008_{0.00}$ & $0.0018_{0.00}$ & $0.0029_{0.01}$ & $0.0084_{0.05}$ & $0.0262_{0.15}$ & $\underline{0.0329}_{0.04}$ & $\textbf{0.0399}_{0.02}$ & $0.0344_{0.04}$ \\
        \cline{2-12}
        
        % ===================== Music4all =====================
        & \multirow{6}*{Music4all}
         & Prec@10  & -- OOM --   & -- OOM -- & $0.0248_{0.00}$ & $0.0334_{0.23}$ & $0.0132_{0.23}$ & $0.0530_{0.80}$ & $\underline{0.0607}_{0.04}$ & $0.0607_{0.05}$ & $\textbf{0.0623}_{0.24}$ \\
         & & Recall@10 & -- OOM --  & -- OOM -- & $0.0052_{0.00}$ & $0.0075_{0.07}$ & $0.0060_{0.05}$ & $0.0150_{0.31}$ & $\textbf{0.0213}_{0.01}$ & $0.0182_{0.04}$ & $0.0183_{0.10}$ \\
         & & NDCG@10  & -- OOM --   & -- OOM -- & $0.0263_{0.00}$ & $0.0357_{0.27}$ & $0.0160_{0.25}$ & $\underline{0.0597}_{0.95}$ & $0.0276_{0.02}$ & $0.0680_{0.04}$ & $\textbf{0.0709}_{0.27}$ \\
         & & Prec@20  & -- OOM --   & -- OOM -- & $0.0224_{0.00}$ & $0.0292_{0.22}$ & $0.0109_{0.19}$ & $0.0456_{0.69}$ & $\underline{0.0527}_{0.02}$ & $0.0513_{0.03}$ & $\textbf{0.0533}_{0.19}$ \\
         & & Recall@20 & -- OOM --  & -- OOM -- & $0.0094_{0.00}$ & $0.0130_{0.11}$ & $0.0086_{0.07}$ & $0.0248_{0.49}$ & $\textbf{0.0342}_{0.02}$ & $0.0294_{0.05}$ & $0.0299_{0.15}$ \\
         & & NDCG@20  & -- OOM --   & -- OOM -- & $0.0248_{0.00}$ & $0.0332_{0.26}$ & $0.0150_{0.22}$ & $\underline{0.0553}_{0.89}$ & $0.0368_{0.03}$ & $0.0625_{0.04}$ & $\textbf{0.0654}_{0.24}$ \\
         \hline
    \end{tabular}}
    \label{exp:noisy}
\end{table*}

\begin{table*}[h]
    \centering
    \caption{Ablation study on vanilla datasets.}
    \begin{tabular}{c|c|c|c|c|c|c|c}
        \hline
        Dataset & Method & Prec@10 & Recall@10 & NDCG@10 & Prec@20 & Recall@20 & NDCG@20 \\
        \hline
        \multicolumn{8}{c}{\nameA{}} \\
        \hline
        \multirow{2}*{ML-100K} & w/o t & 0.2117 & 0.0936 & 0.2254 & 0.1853 & 0.1597 & 0.2265 \\
        ~ & \nameA{} & \textbf{0.2373} & \textbf{0.1026} & \textbf{0.2542} & \textbf{0.2063} & \textbf{0.1746} & \textbf{0.2532} \\
        \hline
        \multirow{2}*{Tmall} & w/o t & 0.0085 & 0.0618 & 0.0364 & 0.0062 & 0.0903 & 0.0441 \\
        ~ & \nameA{} & \textbf{0.0088} & \textbf{0.0643} & \textbf{0.0372} & \textbf{0.0064} & \textbf{0.0936} & \textbf{0.0452} \\
        \hline
        
        \multicolumn{8}{c}{\nameL{}} \\
        \hline
        \multirow{2}*{ML-100K} & w/o t & \textbf{0.2197} & 0.0949 & 0.2652 & 0.1955 & 0.1678 & 0.2692 \\
        ~ & \nameL{} & 0.2188 & \textbf{0.0959} & \textbf{0.2672} & \textbf{0.1974} & \textbf{0.1699} & \textbf{0.2734} \\
        \hline
        \multirow{2}*{Tmall} & w/o t & 0.0058 & 0.0399 & 0.0235 & 0.0046 & 0.0637 & 0.0300 \\
        ~ & \nameL{} & \textbf{0.0078} & \textbf{0.0557} & \textbf{0.0326} & \textbf{0.0059} & \textbf{0.0850} & \textbf{0.0405} \\
        \hline
    \end{tabular}
    \label{tab:ablation:vanilla}
\end{table*}

\begin{table*}[h]
    \centering
    \caption{Ablation study on noisy datasets.}
    \begin{tabular}{c|c|c|c|c|c|c|c}
        \hline
        Dataset & Method & Prec@10 & Recall@10 & NDCG@10 & Prec@20 & Recall@20 & NDCG@20 \\
        \hline
        \multicolumn{8}{c}{\nameA{}} \\
        \hline
        \multirow{2}*{ML-100K} & w/o t & 0.1999 & 0.0876 & 0.2128 & 0.1773 & 0.1529 & 0.2157 \\
        ~ & \nameA{} & \textbf{0.2428} & \textbf{0.1060} & \textbf{0.2607} & \textbf{0.2086} & \textbf{0.1763} & \textbf{0.2576} \\
        \hline
        \multirow{2}*{Tmall} & w/o t & 0.0078 & 0.0552 & 0.0327 & 0.0057 & 0.0808 & 0.0397 \\
        ~ & \nameA{} & \textbf{0.0080} & \textbf{0.0570} & \textbf{0.0331} & \textbf{0.0057} & \textbf{0.0820} & \textbf{0.0399} \\
        \hline

        \multicolumn{8}{c}{\nameL{}} \\
        \hline
        \multirow{2}*{ML-100K} & w/o t & 0.2140 & 0.0944 & 0.2618 & 0.1916 & 0.1665 & 0.2674 \\
        ~ & \nameL{} & \textbf{0.2211} & \textbf{0.0954} & \textbf{0.2666} & \textbf{0.1958} & \textbf{0.1687} & \textbf{0.2706} \\
        \hline
        \multirow{2}*{Tmall} & w/o t & 0.0052 & 0.0352 & 0.0208 & 0.0040 & 0.0548 & 0.0262 \\
        ~ & \nameL{} & \textbf{0.0069} & \textbf{0.0479} & \textbf{0.0281} & \textbf{0.0051} & \textbf{0.0707} & \textbf{0.0344} \\
        \hline
    \end{tabular}
    \label{tab:ablation:noisy}
\end{table*}

\section{Experiments}
In this section, we present the experimental evaluation. The experiments are designed to answer the following research questions:
\begin{itemize}
    \item \textbf{RQ1.} How does the proposed approach perform compared to state-of-the-art denoising and general-purpose recommender systems?
    \item \textbf{RQ2.} To what extent does proposed time-aware embeddings contribute to overall performance?
    \item \textbf{RQ3.} How does our proposed approach successfully downweight noisy edges?
\end{itemize}

\subsection{Experimental Settings}
\subsubsection{Datasets}
To evaluate the performance of \name{}, we conduct experiments on four real-world public datasets. ML-100K \cite{movielens} and ML-1M \cite{movielens} are explicit feedback based datasets, Tmall \cite{tmall} and Music4all \cite{music4all} are implicit feedback based datasets. ML-100K and ML-1M are movie review datasets, while Tmall and Music4all are commercial and music datasets respectively.

\begin{table}[H]
    \centering
    \caption{Data Statistics}
    \begin{tabular}{c|c|c|c|c}
        \hline
        Dataset & \#Users & \#Items & \#Interactions & Density\\
        \hline
        ML-100K & 943 & 1682 & 99,999 & 6.305\% \\
        ML-1M & 6,040 & 3,706 & 1,000,209 & 4.468\% \\
        Tmall & 19030 & 47287 & 98,387 & 0.011\% \\
        Music4all & 16232 & 54152 & 12,310,886 & 1.401\% \\
        \hline
    \end{tabular}
    \label{tab:data}
\end{table}

Data statistics are summarized in Table~\ref{tab:data}. The available timestamps in the datasets are Day-Hour-Minute-Second(ML-100K), Month-Day-Hour-Minute-Second(ML-1M, Tmall, Music4all).

\subsubsection{Baselines}We compare our proposed approach with the state-of-the-art time-aware recommender systems (TIN \cite{zhou2024temporal}, TGSRec\cite{TGSRec}) and denoised recommender systems (T-CE \cite{T-CE}, DeCA \cite{DeCA}, DDRM \cite{ddrm}, BOD \cite{BOD}, PLD \cite{PLD}).
To conduct a more comprehensive analysis, we also compare our proposed approach with 5 general-purpose recommender systems, including 2 sequential (Bert4Rec~\cite{sun2019bert4rec}, CL4SRec~\cite{cl4srec}) and 3 graph-based (LightGCN~\cite{lightgcn}, NCL~\cite{NCL} and SimGCL~\cite{SimGCL}).
Details of baselines in main table are summarized as followed.
\begin{itemize}
    \item TGSRec \cite{TGSRec} uses a temporal collaborative transformer to capture time-aware sequential and collaborative interaction patterns.
    \item TIN \cite{zhou2024temporal} is a time-aware sequential recommendation model that incorporate target-aware temporal encodings.
    \item T-CE \cite{T-CE} denoises recommender systems by using truncated cross entropy loss to remove any positive interactions with cross entropy loss larger than a certain threshold from the training.
    \item DeCA \cite{DeCA} denoises recommender systems by maximizing the agreement between two models as noisy samples tend to induce disagreement between models.
    \item DDRM \cite{ddrm} is a diffusion-based denoised recommender system, which denoises trained user and item embeddings through a diffusion model.
    \item BOD \cite{BOD} denoises recommender systems by reweighting the loss function with a weight generator that is optimized by gradient matching.
    \item PLD \cite{PLD} applies a personalized item resampling method proportional to loss of user-item pairs, motivated by variance in users' personal loss distribution.
\end{itemize}

\subsubsection{Implementation}To compare the performance of proposed approaches for prediction and debiasing, we conduct experiments on both vanilla datasets and noisy datasets. Only records in 2013 with corresponding user/item having more than or equal to 10 records are kept in Tmall dataset. Only records in 2020 with corresponding user/item having more than or equal to 5 records are kept in Music4all dataset. Datasets are sequentially split into training set and test set by 7:3.
Noisy datasets are constructed by randomly adding 20\% noisy interactions into the training set of preprocessed datasets, while the test set remains the same. For each user, the earliest timestamp of interaction with this user in test set is considered as query time for testing. All results are taken average of 4 rounds of experiments with different random seeds.  All baselines in the main table are run with their official repositories, additional general-purpose baselines are run with the implementation in \cite{selfrec}. Our codes are available at \href{https://anonymous.4open.science/r/TintCF-6D9F}{https://anonymous.4open.science/r/TintCF-6D9F}.

\subsubsection{Parameter Settings}. We conduct grid search on batch size $[64, 256, 1024, 2048]$ and learning rate $[1e^{-4}, 5e^{-4}, 1e^{-3}]$. For NCL, the number of clusters are searched in $[100, 200, 1000, 2000]$. NeuMF \cite{NeuMF} is selected as the base model for both T-CE and DeCA as it has the best performance reported in both papers. Similarly, SGL is selected for DDRM, and LightGCN is selected for PLD. For PLD, we iterate over $[0, 0.1, 0.2, 0.3, 0.4, 0.5]$ as noise levels and report the one with best performance.\\

\subsection{Main Results (RQ1)}
Experimental results on vanilla datasets are shown in Table \ref{exp:vanilla} and experimental results on noisy datasets are shown in Table \ref{exp:noisy}. Experimental results for additional general-purpose baselines are reported in Table.~\ref{exp:additional}.

\nameA{} and \nameL{} significantly outperforms all other baselines on implicit feedback datasets (Tmall and Music4all). \nameL{}, which generally follows the training scheme of BOD \cite{BOD}, outperforms BOD. This confirms the ability of our time-aware embeddings for boosting models' denoising ability. The relative average improvements in percentage against the 8 baselines per metric %on average
on Tmall and Music4all datasets are shown in Table~\ref{tab:improvement}, reaching an improvement of 12.88\% on average across all metrics. Notably, our approaches reach almost 20\% performance increase in NDCG@20 on noisy datasets. This confirms that time information can indeed help the models to focus on users' primary interests and recommend them in the top of lists. Moreover, improvements on noisy implicit feedback datasets are generally higher than the improvements on vanilla implicit feedback datasets, which further confirms the denoising ability of time information.

In implicit feedback datasets, timestamps are promptly recorded along with user actions. On the contrary, in explicit feedback based datasets, timestamps do not necessarily indicate the actual time that interaction takes place. However, even with those noisy timestamps in explicit feedback based datasets (ML-100K and ML-1M), \nameA{} and \nameL{} still reach better performance compared to other baselines, demonstrating the effectiveness of \name{}.

\begin{table}[H]
    \centering
    \caption{Relative improvement by percentage on implicit feedback datasets.}
    \begin{tabular}{c|c|c|c|c|c|c}
        \hline
        ~ & \multicolumn{3}{c|}{Vanilla} & \multicolumn{3}{c}{Noisy} \\
        \hline
        Metric & Prec & Recall & NDCG & Prec & Recall & NDCG \\
        \hline
        @10 & 12.99 & 5.77 & 13.76 & 16.91 & 10.99 & 16.51 \\
        @20 & 13.82 & 2.21 & 17.65 & 16.85 & 7.28 & 19.77 \\
        \hline
    \end{tabular}
    \label{tab:improvement}
\end{table}

\begin{figure*}[t]
    \centering
    \begin{minipage}{0.24\textwidth}
        \centering
        \includegraphics[width=\linewidth]{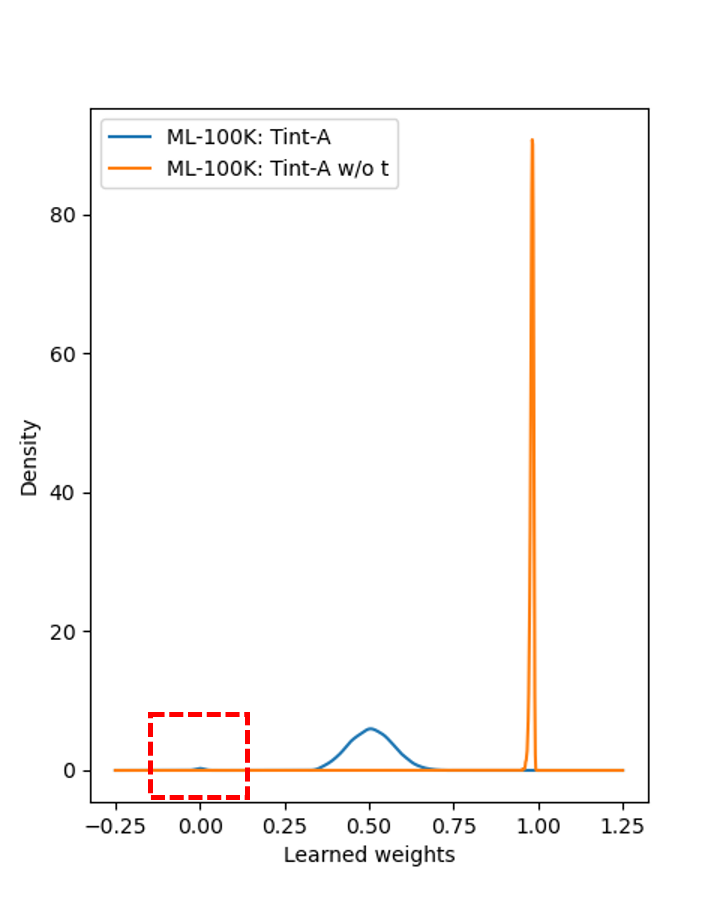}
    \end{minipage}%
    \begin{minipage}{0.24\textwidth}
        \centering
        \includegraphics[width=\linewidth]{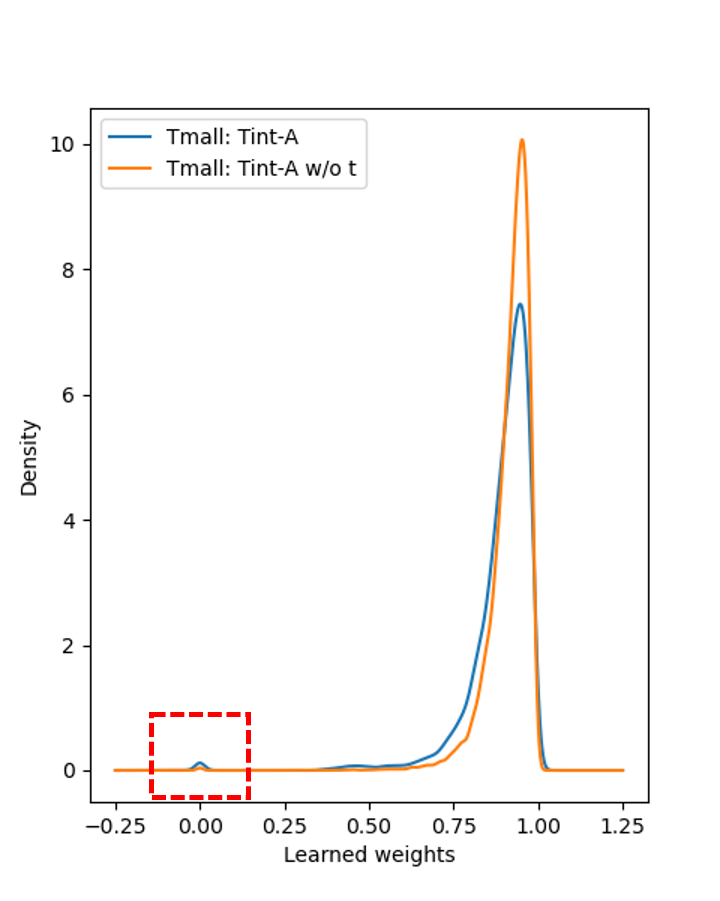}
    \end{minipage}
    \begin{minipage}{0.24\textwidth}
        \centering
        \includegraphics[width=\linewidth]{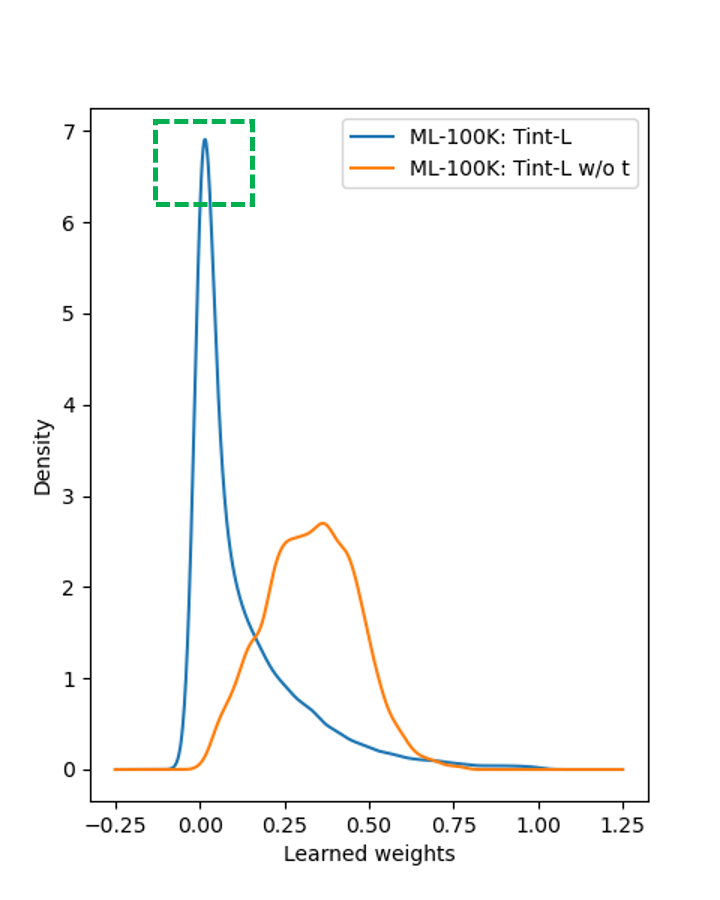}
    \end{minipage}%
    \begin{minipage}{0.24\textwidth}
        \centering
        \includegraphics[width=\linewidth]{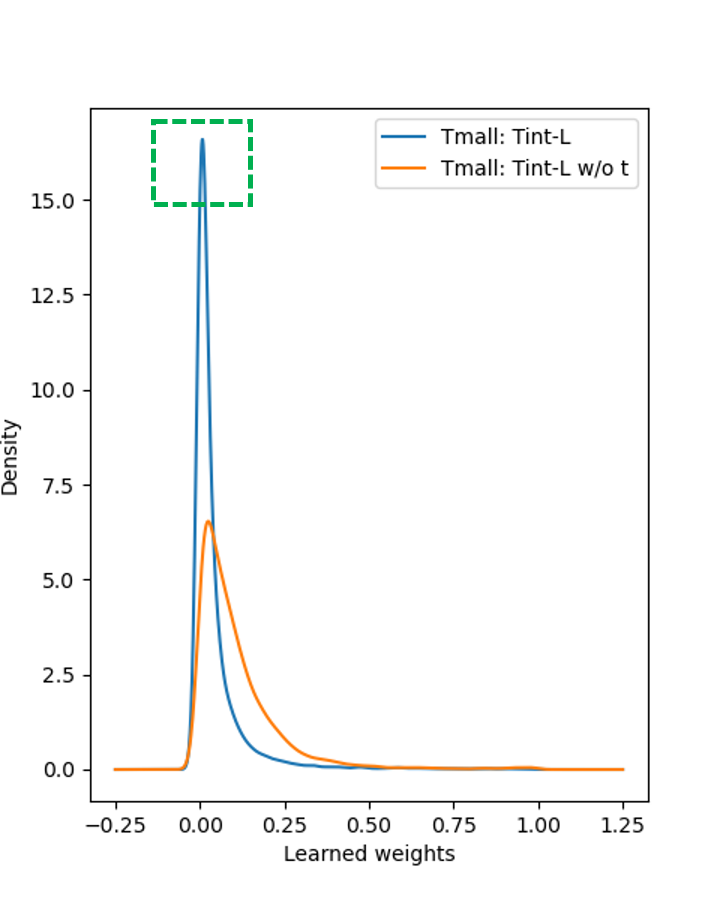}
    \end{minipage}
    \caption{Visualization of learned weights of noisy edges. Left to right: Tint-A on ML-100K, Tint-A on Tmall, Tint-L on ML-100K, Tint-L on Tmall. Best viewed in color.}
    \label{fig:vis:w}
\end{figure*}
\subsection{Ablation Studies (RQ2)}
We further conduct ablation studies on ML-100K and Tmall datasets to evaluate the actual contribution of time-aware embeddings on the overall performance. We train \nameA{} and \nameL{} by removing time-aware embeddings from reliability score function or weight generator. Specifically, removing time-aware embeddings from \nameL{} will degrade to BOD. Experimental results on vanilla datasets and noisy datasets are shown in Table \ref{tab:ablation:vanilla} and Table \ref{tab:ablation:noisy} respectively. We can observe that removing time-aware embeddings causes a performance drop of both \nameA{} and \nameL{} across all datasets. Therefore time-aware embeddings indeed effectively improves the model performance on both vanilla and noisy datasets. 

\subsection{Effect of Time-aware embeddings (RQ3)}
To further demonstrate that time-aware embeddings can effectively downweight noisy edges, we visualize the reliability scores and weights of noisy data samples on noisy Tmall and ML-100K datasets. Density distributions of learned weights of noisy data samples are estimated using Gaussian kernel density estimation \cite{scott2015multivariate}. Results are shown in Fig.~\ref{fig:vis:w}. We can observe that density distributions trained with time information (blue lines) are more skewed to the left compared with removing time information (orange lines) in all cases, meaning noisy data samples are getting a lower weight with time information involved. This confirms the denoising ability of time-aware embeddings. Also, as we allow hard assignment in \nameA{}, more noisy edges are directly pruned with time information involved as shown in red boxes. Although \nameL{} does not allow hard pruning, the distribution peaks near zero-weight in green boxes.

\begin{table*}[!htpb]
    \centering
    \caption{Additional experiments on vanilla and noisy datasets for general-propose baselines. Stds(\%) are in footnotes.}
    \resizebox{\textwidth}{!}{
    \begin{tabular}{c|c|ccccc|ccccc}
        \hline
        \multirow{2}{*}{Dataset} & \multirow{2}{*}{Metric}
        & \multicolumn{5}{c|}{Vanilla Dataset}
        & \multicolumn{5}{c}{Noisy Dataset} \\
        \cline{3-12}
        & 
        & Bert4Rec & CL4SRec & LightGCN & NCL & SimGCL
        & Bert4Rec & CL4SRec & LightGCN & NCL & SimGCL \\
        \hline

        \multirow{6}{*}{ML-100K}
        & Prec@10
        & $0.0040_{0.04}$ & $0.0033_{0.02}$ & $0.2333_{0.73}$ & $0.2349_{0.81}$ & $0.2321_{0.42}$
        & $0.0024_{0.03}$ & $0.0029_{0.02}$ & $0.2195_{0.49}$ & $0.2158_{0.87}$ & $0.2232_{0.62}$ \\
        & Recall@10
        & $0.0398_{0.36}$ & $0.0329_{0.20}$ & $0.1038_{0.38}$ & $0.1011_{0.54}$ & $0.1028_{0.23}$
        & $0.0244_{0.29}$ & $0.0289_{0.20}$ & $0.0960_{0.23}$ & $0.0950_{0.42}$ & $0.0977_{0.49}$ \\
        & NDCG@10
        & $0.0184_{0.11}$ & $0.0150_{0.13}$ & $0.2508_{0.83}$ & $0.2540_{0.80}$ & $0.2481_{0.51}$
        & $0.0113_{0.16}$ & $0.0131_{0.12}$ & $0.2390_{0.38}$ & $0.2329_{1.00}$ & $0.2408_{0.60}$ \\
        & Prec@20
        & $0.0034_{0.04}$ & $0.0031_{0.00}$ & $0.2018_{0.30}$ & $0.2023_{0.51}$ & $0.2007_{0.36}$
        & $0.0023_{0.03}$ & $0.0026_{0.02}$ & $0.1883_{0.41}$ & $0.1870_{0.72}$ & $0.1951_{0.38}$ \\
        & Recall@20
        & $0.0687_{0.75}$ & $0.0610_{0.05}$ & $0.1734_{0.16}$ & $0.1692_{0.63}$ & $0.1716_{0.18}$
        & $0.0451_{0.62}$ & $0.0512_{0.48}$ & $0.1596_{0.37}$ & $0.1594_{0.53}$ & $0.1664_{0.51}$ \\
        & NDCG@20
        & $0.0256_{0.20}$ & $0.0220_{0.13}$ & $0.2491_{0.54}$ & $0.2501_{0.68}$ & $0.2469_{0.41}$
        & $0.0165_{0.21}$ & $0.0186_{0.13}$ & $0.2347_{0.38}$ & $0.2309_{0.92}$ & $0.2399_{0.54}$ \\
        \hline

        \multirow{6}{*}{ML-1M}
        & Prec@10
        & $0.0043_{0.15}$ & $0.0011_{0.03}$ & $0.1574_{0.03}$ & $0.1582_{0.05}$ & $0.1574_{0.18}$
        & $0.0028_{0.05}$ & $0.0016_{0.06}$ & $0.1566_{0.04}$ & $0.1582_{0.08}$ & $0.1567_{0.09}$ \\
        & Recall@10
        & $0.0426_{1.55}$ & $0.0113_{0.31}$ & $0.0393_{0.02}$ & $0.0391_{0.02}$ & $0.0388_{0.05}$
        & $0.0279_{0.47}$ & $0.0159_{0.62}$ & $0.0387_{0.03}$ & $0.0393_{0.08}$ & $0.0388_{0.04}$ \\
        & NDCG@10
        & $0.0162_{0.67}$ & $0.0052_{0.17}$ & $0.1658_{0.04}$ & $0.1664_{0.02}$ & $0.1651_{0.22}$
        & $0.0094_{0.15}$ & $0.0067_{0.27}$ & $0.1648_{0.02}$ & $0.1665_{0.13}$ & $0.1650_{0.13}$ \\
        & Prec@20
        & $0.0030_{0.04}$ & $0.0011_{0.03}$ & $0.1427_{0.07}$ & $0.1429_{0.09}$ & $0.1423_{0.12}$
        & $0.0024_{0.03}$ & $0.0015_{0.07}$ & $0.1426_{0.06}$ & $0.1432_{0.13}$ & $0.1425_{0.14}$ \\
        & Recall@20
        & $0.0598_{0.74}$ & $0.0216_{0.63}$ & $0.0708_{0.09}$ & $0.0704_{0.13}$ & $0.0705_{0.08}$
        & $0.0481_{0.50}$ & $0.0297_{1.40}$ & $0.0711_{0.02}$ & $0.0710_{0.12}$ & $0.0709_{0.11}$ \\
        & NDCG@20
        & $0.0205_{0.47}$ & $0.0078_{0.24}$ & $0.1591_{0.05}$ & $0.1593_{0.09}$ & $0.1583_{0.15}$
        & $0.0142_{0.10}$ & $0.0102_{0.46}$ & $0.1588_{0.04}$ & $0.1596_{0.16}$ & $0.1587_{0.17}$ \\
        \hline

        \multirow{6}{*}{Tmall}
        & Prec@10
        & $0.0033_{0.01}$ & $0.0006_{0.04}$ & $0.0065_{0.01}$ & $0.0048_{0.01}$ & $0.0060_{0.01}$
        & $0.0034_{0.01}$ & $0.0039_{0.06}$ & $0.0056_{0.00}$ & $0.0045_{0.01}$ & $0.0053_{0.01}$ \\
        & Recall@10
        & $0.0328_{0.07}$ & $0.0062_{0.36}$ & $0.0453_{0.06}$ & $0.0338_{0.10}$ & $0.0411_{0.06}$
        & $0.0339_{0.10}$ & $0.0392_{0.56}$ & $0.0385_{0.05}$ & $0.0308_{0.10}$ & $0.0352_{0.05}$ \\
        & NDCG@10
        & $0.0176_{0.03}$ & $0.0030_{0.14}$ & $0.0264_{0.04}$ & $0.0193_{0.05}$ & $0.0248_{0.04}$
        & $0.0187_{0.07}$ & $0.0256_{0.32}$ & $0.0227_{0.05}$ & $0.0177_{0.07}$ & $0.0214_{0.03}$ \\
        & Prec@20
        & $0.0023_{0.00}$ & $0.0005_{0.03}$ & $0.0050_{0.01}$ & $0.0037_{0.01}$ & $0.0045_{0.01}$
        & $0.0023_{0.01}$ & $0.0024_{0.04}$ & $0.0043_{0.01}$ & $0.0034_{0.01}$ & $0.0039_{0.00}$ \\
        & Recall@20
        & $0.0452_{0.10}$ & $0.0099_{0.62}$ & $0.0709_{0.12}$ & $0.0526_{0.09}$ & $0.0615_{0.13}$
        & $0.0459_{0.12}$ & $0.0488_{0.74}$ & $0.0592_{0.10}$ & $0.0470_{0.15}$ & $0.0521_{0.06}$ \\
        & NDCG@20
        & $0.0208_{0.03}$ & $0.0039_{0.21}$ & $0.0333_{0.05}$ & $0.0244_{0.05}$ & $0.0304_{0.07}$
        & $0.0218_{0.08}$ & $0.0280_{0.36}$ & $0.0283_{0.07}$ & $0.0221_{0.07}$ & $0.0261_{0.03}$ \\
        \hline

        \multirow{6}{*}{Music4all}
        & Prec@10
        & $0.0043_{0.02}$ & $0.0012_{0.00}$ & $0.0301_{0.24}$ & $0.0178_{0.13}$ & $0.0305_{0.43}$
        & $0.0037_{0.01}$ & $0.0004_{0.00}$ & $0.0252_{0.50}$ & $0.0161_{0.04}$ & $0.0288_{0.54}$ \\
        & Recall@10
        & $0.0435_{0.21}$ & $0.0121_{0.02}$ & $0.0070_{0.06}$ & $0.0038_{0.06}$ & $0.0074_{0.14}$
        & $0.0367_{0.05}$ & $0.0043_{0.02}$ & $0.0057_{0.14}$ & $0.0032_{0.01}$ & $0.0068_{0.16}$ \\
        & NDCG@10
        & $0.0221_{0.09}$ & $0.0087_{0.03}$ & $0.0322_{0.26}$ & $0.0189_{0.14}$ & $0.0328_{0.52}$
        & $0.0184_{0.04}$ & $0.0034_{0.01}$ & $0.0268_{0.57}$ & $0.0172_{0.03}$ & $0.0309_{0.62}$ \\
        & Prec@20
        & $0.0033_{0.01}$ & $0.0007_{0.00}$ & $0.0264_{0.24}$ & $0.0160_{0.12}$ & $0.0271_{0.39}$
        & $0.0029_{0.01}$ & $0.0002_{0.00}$ & $0.0226_{0.44}$ & $0.0146_{0.01}$ & $0.0255_{0.46}$ \\
        & Recall@20
        & $0.0668_{0.25}$ & $0.0145_{0.03}$ & $0.0122_{0.13}$ & $0.0070_{0.09}$ & $0.0127_{0.22}$
        & $0.0575_{0.11}$ & $0.0049_{0.02}$ & $0.0102_{0.24}$ & $0.0060_{0.01}$ & $0.0119_{0.26}$ \\
        & NDCG@20
        & $0.0279_{0.10}$ & $0.0093_{0.03}$ & $0.0300_{0.27}$ & $0.0179_{0.15}$ & $0.0309_{0.48}$
        & $0.0236_{0.04}$ & $0.0035_{0.01}$ & $0.0253_{0.53}$ & $0.0162_{0.02}$ & $0.0290_{0.56}$ \\
        \hline
    \end{tabular}
    }
    \label{exp:additional}
\end{table*}
\section{Related Works}
\noindent\textbf{Denoising Recommender Systems}. 
The goal of denoising recommender systems is to identify noisy edges (i.e., interactions) according to different principles and purify the interactions in different learning phases. For example, \cite{RocSE,RGCF} determine interactions that connect dissimilar users and items as noisy and downweight those noisy edges in the adjacency matrix. \cite{T-CE} observes that noisy interaction tends to have larger losses, and prunes those positive samples with larger losses in the training stage. \cite{SGDL} instead memorizes clean samples in the early training stage, and reweights the objective function by the similarity between samples and those memorized clean samples. Another group of work achieves robustness by agreements. \cite{BOD} trains a weight generator that maximizes the agreement between gradients of different objective functions, then the weights are applied to reweight objective functions to train backbone models. \cite{DeCA} maximizes  the agreement between different models as they observe that different models make relatively similar predictions for clean samples. \cite{ddrm} focuses on denoising diffusion recommender models by inserting noises in the forward process. \cite{10771708} denoises social information in social recommender systems from two perspectives, noises within user-user relationships and noises induced by the heterogeneous meta-paths, e.g., user-user-item.
A new line of work focuses on diffusion based recommender systems \cite{ddrm} which denoise learned embeddings by passing them through the forward and backward process of diffusion models.
There also exist some works that are not explicitly designed for denoising recommender systems but potentially increase the model robustness. \cite{SGL} generates different views of graphs through graph structure augmentation, and maximizes the agreement between different views through contrastive learning. \cite{SimGCL} instead augments  user and item embeddings by adding perturbation to GNN layers, and maximizes  agreement between different views of embeddings.

\noindent\textbf{Time-aware Recommender Systems}. Temporal order has been widely applied in sequential recommendation \cite{kang2018self,10.1145/3523227.3546785,sun2019bert4rec}. Lately, several works investigate the possibility of directly incorporating timestamps into sequential models \cite{10.1145/3604915.3608830,TGSRec,kazemi2019time2vec, 10.1145/3605356}. \cite{kazemi2019time2vec} proposes Time2vec, a representation of time based on periodic activation function that can be used by many models. \cite{10.1145/3604915.3608830} considers time sequence as a multi-dimensional time-series in sequential recommendation, and uses kernel-based and embedding-based approaches for time representation learning. \cite{TGSRec} proposes a temporal collaborative transformer layer to exploit the temporal collaborative signals in evolving user-item preferences.
\cite{10.1145/3605356} introduces a time-aware sequential model to capture the user shifting interest, where a decaying function with time intervals as input is adopted to control the update of hidden states over time. \cite{zhou2024temporal} studies the semantic-temporal correlation between behaviors and target by adopting a target-aware temporal encoding. \cite{TGSRec} models recommendation as a continuous-time user–item graph and uses a temporal collaborative transformer to capture time-aware sequential and collaborative interaction patterns.
However, limited attempts have been made to incorporate time information in the context of graph-based collaborative filtering and denoising recommender systems. \cite{9378444} formulates the temporal order of interactions into an item-item graph and leverages temporal information from three different views: user-user, user-item and item-item.

\section{Conclusion}
In this paper, we propose a simple yet effective mechanism, \name{}, for generating time-aware user/item embeddings, which integrates %is defined as the addition of 
user/item embeddings with timestamp embeddings.
To leverage this mechanism for denoising temporal bipartite graphs in recommender systems, we apply time-aware embeddings in two strategies: reweighting the adjacency matrix (\nameA{}) and reweighting the loss function (\nameL{}). \nameA{} reweights the adjacency matrix through a cosine based reliability score function, while \nameL{} reweights the loss function through a weight generator. Both the reliability score function and the weight generator take time-aware embeddings as input. Extensive experiments on four real-world datasets demonstrate the efficacy of our proposed approach. The mechanism we propose for generating time-aware embeddings is not limited to these two strategies, but could be easily integrated with other neural graph collaborative filtering frameworks as well. Moving forward, future works will focus on investigating other time-aware neural graph collaborative filtering algorithms by integrating our method.
We will also consider expanding the proposed method to denoise not only user-item interactions but also noise in user profiles or item attributes.

\clearpage
\section*{Acknowledgment}
This work is supported by NSF (2433308).
The content of the information in this document does not necessarily reflect the position or the policy of the Government, and no official endorsement should be inferred.  The U.S. Government is authorized to reproduce and distribute reprints for Government purposes notwithstanding any copyright notation here on.

\end{document}